\documentclass[11pt,reqno]{amsart}
\usepackage{amscd,amssymb,amsmath,amsthm}
\usepackage[arrow,matrix]{xy}
\usepackage{graphicx}
\usepackage{cite}
\newtheorem{thm}[subsection]{Theorem}

\newtheorem{pro}[subsection]{Proposition}

\newtheorem{defn}[subsection]{Definition}

\numberwithin{equation}{section} 

\newcommand{\s}{{\sigma}}
\newcommand{\de}{{\xi}}

\newcommand{\bea}{\begin{eqnarray}}
\newcommand{\eea}{\end{eqnarray}}

\begin{document}
\title[Phase Transitions for a model]{Phase Transitions for a model with uncountable set of spin values on a Cayley tree}

\author{Yu. Kh. Eshkabilov, U. A. Rozikov, G.I. Botirov}

 \address{Yu.\ Kh.\ Eshkabilov\\ Faculty of Mechanics and Mathematics National University of Uzbekistan,
Tashkent, Uzbekistan.} \email {yusup62@mail.ru}

\address{U.\ A.\ Rozikov\\ Institute of mathematics, National University of Uzbekistan,
Tashkent, Uzbekistan.} \email {rozikovu@yandex.ru}

\address{G.\ I.\ Botirov\\ Faculty of Physics and Mathematics of Bukhara State University,
Bukhara, Uzbekistan.} \email {botirovg@yandex.ru}
\begin{abstract} In this paper we consider a model with nearest-neighbor
interactions and with the set $[0,1]$ of spin values, on a Cayley
tree of order $k \geq 2$. To study translation-invariant Gibbs
measures of the model we drive an nonlinear functional equation.
For $k=2$ and 3 under some conditions on parameters of the model
we prove non-uniqueness of translation-invariant Gibbs measures
(i.e. there are phase transitions).
\end{abstract}
\maketitle



\section{Introduction} \label{sec:intro}

A central problem in the theory of Gibbs measures is to describe
infinite-volume (or limiting) Gibbs measures corresponding to a
given Hamiltonian.

In order to study the phase transition (Gibbs measures) problem
for a system on $Z^d$ and on Cayley tree there are two different
methods: Pirogov-Sinai theory on $Z^d$, Markov random field theory
and recurrent equations of this theory on Cayley tree.

The papers \cite{2}-\cite{4},\cite{p1a}, \cite{11}-\cite{p2},
\cite{13}, \cite{14}, \cite{16} are devoted to models with a {\it
finite} set of spin values. It were shown that these models have
finitely many translation-invariant  and uncountable numbers of
the non-translation-invariant extreme Gibbs measures. Also for
several models (see, for example, \cite{p1, p1a, p2}) it were
proved that there exist three  periodic  Gibbs  measures (which
are invariant with respect to normal  subgroups  of  finite index
of the group representation of the Cayley tree) and there are
uncountable number of non-periodic Gibbs measures.

In \cite{6} the Potts model with a {\it countable} set of spin
values on a Cayley tree is considered and it was showed that the
set of translation-invariant splitting Gibbs measures of the model
contains at most one point, independently on parameters of the
Potts model with countable set of spin values on the Cayley tree.
This is a crucial difference from the models with a finite set of
spin values, since the last ones may have more than one
translation-invariant Gibbs measures.

This paper is continuation of our investigations \cite{ehr},
\cite{ehr1}, \cite{re}. In \cite{re} models (Hamiltonians) with
nearest-neighbor interactions and with the ({\it uncountable}) set
$[0,1]$ of spin values, on a Cayley tree $\Gamma^k$ of order
$k\geq 1$ were studied.

We reduced the problem  to the description of the solutions of
some nonlinear integral equation. Then for $k = 1$ we showed that
the integral equation has a unique solution. In case $k\geq 2$
some models (with the set $[0, 1]$ of spin values) which have a
unique splitting Gibbs measure are constructed.  In our next paper
\cite{ehr} it was found a sufficient condition on Hamiltonian of
the model with an uncountable set of spin values under which the
model has unique translation-invariant splitting Gibbs measure. In
\cite{ehr1} we have constructed several examples of models with
uncountable set of spin values which have phase transitions.In
this paper we shall construct new models with nearest-neighbor
interactions and with the set $[0,1]$ of spin values, on a Cayley
tree order $k$. We prove that each of the constructed model has a
phase transition.

\section{Preliminaries}

A Cayley tree $\Gamma^k=(V,L)$ of order $k\geq 1$ is an infinite
homogeneous tree, i.e., a graph without cycles, with exactly $k+1$
edges incident to each vertices. Here $V$ is the set of vertices
and $L$ that of edges (arcs).

Consider models where the spin takes values in the set $[0,1]$,
and is assigned to the vertexes of the tree. For $A\subset V$ a
configuration $\s_A$ on $A$ is an arbitrary function $\s_A:A\to
[0,1]$. Denote $\Omega_A=[0,1]^A$ the set of all configurations on
$A$. A configuration $\sigma$ on $V$ is then defined as a function
$x\in V\mapsto\sigma (x)\in [0,1]$; the set of all configurations
is $[0,1]^V$. We consider the (formal) Hamiltonian of the model is
:
\begin{equation}\label{e1.1}
 H=H_{\theta,\beta}(\sigma)=-\sum_{\langle x,y\rangle\in L}
\de_{\sigma(x)\sigma(y)}(\theta,\beta), \ \ \ \sigma\in\Omega_A
\end{equation}
where $\theta \in R$ is a coupling constant, $\beta={1\over T}$,
$T>0 $ is temperature and $\de: (u,v)\in [0,1]^2\to \de_{uv}\in R$
is a given bounded, measurable function. As usually, $\langle
x,y\rangle$ stands for nearest neighbor vertices.

 Note that, a Gibbs measures for the model (2.1) with
$\de_{t,u}(J,\beta):=J\de_{t,u}$ are investigated in \cite{ehr},
\cite{ehr1}, \cite{re}.

Let $\lambda$ be the Lebesgue measure on $[0,1]$. On the set of
all configurations on $A$ the a priori measure $\lambda_A$ is
introduced as the $|A|$fold product of the measure $\lambda$. Here
and further on $|A|$ denotes the cardinality of $A$.   We consider
a standard sigma-algebra ${\mathcal B}$ of subsets of
$\Omega=[0,1]^V$ generated by the measurable cylinder subsets.
 A probability measure $\mu$ on $(\Omega,{\mathcal B})$
is called a Gibbs measure (with Hamiltonian $H$) if it satisfies
the DLR equation, namely for any $n=1,2,\ldots$ and
$\sigma_n\in\Omega_{V_n}$:
$$\mu\left(\left\{\sigma\in\Omega :\;
\sigma\big|_{V_n}=\sigma_n\right\}\right)= \int_{\Omega}\mu ({\rm
d}\omega)\nu^{V_n}_{\omega|_{W_{n+1}}} (\sigma_n),$$ where
$\nu^{V_n}_{\omega|_{W_{n+1}}}$ is the conditional Gibbs density
$$ \nu^{V_n}_{\omega|_{W_{n+1}}}(\sigma_n)=\frac{1}{Z_n\left(
\omega\big|_{W_{n+1}}\right)}\exp\;\left(-\beta H
\left(\sigma_n\,||\,\omega\big|_{W_{n+1}}\right)\right).
$$
Here and below, $W_l$ stands for a `sphere' and $V_l$ for a `ball'
on the tree, of radius $l=1,2,\ldots$, centered at a fixed vertex
$x^0$ (an origin):
$$W_l=\{x\in V: d(x,x^0)=l\},\;\;V_l=\{x\in V: d(x,x^0)\leq l\};$$
and
$$L_n=\{\langle x,y\rangle\in L: x,y\in V_n\};$$
distance $d(x,y)$, $x,y\in V$, is the length of (i.e. the number
of edges in) the shortest path connecting $x$ with $y$.
$\Omega_{V_n}$ is the set of configurations in $V_n$ (and
$\Omega_{W_n}$ that in $W_n$; see below). Furthermore,
$\sigma\big|_{V_n}$ and $\omega\big|_{W_{n+1}}$ denote the
restrictions of configurations $\sigma,\omega\in\Omega$ to $V_n$
and $W_{n+1}$, respectively. Next, $\sigma_n:\;x\in V_n\mapsto
\sigma_n(x)$ is a configuration in $V_n$ and
$H\left(\sigma_n\,||\,\omega\big|_{W_{n+1}}\right)$ is defined as
the sum $H\left(\sigma_n\right)+U\left(\sigma_n,
\omega\big|_{W_{n+1}}\right)$ where
$$H\left(\sigma_n\right)
=-\sum_{\langle x,y\rangle\in L_n}\de_{\sigma_n(x)\sigma_n(y)},$$
$$U\left(\sigma_n,
\omega\big|_{W_{n+1}}\right)= -\sum_{\langle x,y\rangle:\;x\in
V_n, y\in W_{n+1}} \de_{\sigma_n(x)\omega (y)}.$$ Finally,
$Z_n\left(\omega\big|_{W_{n+1}}\right)$ stands for the partition
function in $V_n$, with the boundary condition
$\omega\big|_{W_{n+1}}$:
$$Z_n\left(\omega\big|_{W_{n+1}}\right)=
\int_{\Omega_{V_n}} \exp\;\left(-\beta H
\left({\widetilde\sigma}_n\,||\,\omega
\big|_{W_{n+1}}\right)\right)\lambda_{V_n}(d{\widetilde\sigma}_n).$$

Due to the nearest-neighbor character of the interaction, the
Gibbs measure possesses a natural Markov property: for given a
configuration $\omega_n$ on $W_n$, random configurations in
$V_{n-1}$ (i.e., `inside' $W_n$) and in $V\setminus V_{n+1}$
(i.e., `outside' $W_n$) are conditionally independent.

We use a standard definition of a translation-invariant measure
(see, e.g., \cite{12}).
 The main object of study in this
paper are translation-invariant Gibbs measures for the model
(\ref{e1.1}) on Cayley tree.

Write $x<y$ if the path from $x^0$ to $y$ goes through $x$. Call
vertex $y$ a direct successor of $x$ if $y>x$ and $x,y$ are
nearest neighbors. Denote by $S(x)$ the set of direct successors
of $x$. Observe that any vertex $x\ne x^0$ has $k$ direct
successors and $x^0$ has $k+1$.

Let $h:\;x\in V\mapsto h_x=(h_{t,x}, t\in [0,1]) \in R^{[0,1]}$ be
mapping of $x\in V\setminus \{x^0\}$.  Given $n=1,2,\ldots$,
consider the probability distribution $\mu^{(n)}$ on
$\Omega_{V_n}$ defined by
\begin{equation}\label{e2}
\mu^{(n)}(\sigma_n)=Z_n^{-1}\exp\left(-\beta H(\sigma_n)
+\sum_{x\in W_n}h_{\sigma(x),x}\right),
\end{equation}
 Here, as before, $\sigma_n:x\in V_n\mapsto
\sigma(x)$ and $Z_n$ is the corresponding partition function:
\begin{equation}\label{e3} Z_n=\int_{\Omega_{V_n}}
\exp\left(-\beta H({\widetilde\sigma}_n) +\sum_{x\in
W_n}h_{{\widetilde\sigma}(x),x}\right)
\lambda_{V_n}({d\widetilde\s_n}).
\end{equation}

The probability distributions $\mu^{(n)}$ are compatible if for
any $n\geq 1$ and $\sigma_{n-1}\in\Omega_{V_{n-1}}$:
\begin{equation}\label{e4}
\int_{\Omega_{W_n}}\mu^{(n)}(\sigma_{n-1}\vee\omega_n)\lambda_{W_n}(d(\omega_n))=
\mu^{(n-1)}(\sigma_{n-1}).
\end{equation} Here
$\sigma_{n-1}\vee\omega_n\in\Omega_{V_n}$ is the concatenation of
$\sigma_{n-1}$ and $\omega_n$. In this case there exists a unique
measure $\mu$ on $\Omega_V$ such that, for any $n$ and
$\sigma_n\in\Omega_{V_n}$, $\mu \left(\left\{\sigma
\Big|_{V_n}=\sigma_n\right\}\right)=\mu^{(n)}(\sigma_n)$.

\begin{defn} The measure $\mu$ is called {\it splitting
Gibbs measure} corresponding to Hamiltonian (\ref{e1.1}) and
function $x\mapsto h_x$, $x\neq x^0$. \end{defn}

The following statement describes conditions on $h_x$ guaranteeing
compatibility of the corresponding distributions
$\mu^{(n)}(\sigma_n).$

 \begin{pro}\label{p1}\cite{re} {\it The probability distributions
$\mu^{(n)}(\sigma_n)$, $n=1,2,\ldots$, in} (\ref{e2}) {\sl are
compatible iff for any $x\in V\setminus\{x^0\}$ the following
equation holds:
\begin{equation}\label{e5}
 f(t,x)=\prod_{y\in S(x)}{\int_0^1\exp(\beta\de_{tu})f(u,y)du \over \int_0^1\exp(\beta{\de_{0u}})f(u,y)du}.
 \end{equation}
Here, and below  $f(t,x)=\exp(h_{t,x}-h_{0,x}), \ t\in [0,1]$ and
$du=\lambda(du)$ is the Lebesgue measure.}
\end{pro}

From Proposition 2.2 it follows that for any $h=\{h_x\in
R^{[0,1]},\ \ x\in V\}$ satisfying (\ref{e5}) there exists a
unique Gibbs measure $\mu$ and vice versa. However, the analysis
of solutions to (\ref{e5}) is not easy. This difficulty depends on
the given function $\xi$.

Let $\xi_{tu}$ is a continuous function and we are going to
construct functions $\xi_{tu}$ under which the equation (\ref{e5})
has at least two solutions in the class of translational-invariant
functions $f(t,x)$, i.e $f(t,x)=f(t),$ for any $x\in V$. For such
functions equation (\ref{e5}) can be written as
\begin{equation}\label{e1.2}
f(t)=\left({\int_0^1K(t,u)f(u)du\over \int_0^1
K(0,u)f(u)du}\right)^k,
\end{equation}
where $K(t,u)=\exp(\beta \xi_{tu}), f(t)>0, t,u\in [0,1].$

We put
$$C^+[0,1]=\{f\in C[0,1]: f(x)\geq 0\}.$$
We are interested to positive continuous solutions to
(\ref{e1.2}).

\section{A representation of solutions}

For every $k\in\mathbb{N}$ we consider an integral operator
$H_{k}$ acting in the cone $C^{+}[0,1]$ as
$$(H_{k}f)(t)=\int^{1}_{0}K(t,u)f^{k}(u)du, \,\ k\in\mathbb{N}.$$

The operator $H_{k}$ is called Hammerstein's integral operator of
order $k$. Clearly that, if $k\geq2$ then $H_{k}$ is a nonlinear
operator.

It is known that the set of translational invariant Gibbs measures
of the model (2.1) is described by the fixed points of the
Hammerstein's operator (\cite{re}).

Let $k \geq 2$ in the model (2.1) and
$$\xi_{t,u}=\xi_{t,u}(\theta, \beta)=\frac{1}{\beta} \ln \left(1+\theta
\sqrt[3]{4(t-\frac{1}{2})(u-\frac{1}{2})}\right), \ \ t, u \in
[0,1]$$ where $0 \leq \theta <1$. Then for the kernel $K(t,u)$ of
the Hammerstein's operator $H_k$ we have $$K(t,u)=1+\theta
\sqrt[3]{4(t-\frac{1}{2})(u-\frac{1}{2})}.$$

We define the operator $V_k:(x,y) \in R^2 \rightarrow (C_1,C_2)
\in R^2$ by
$$V_k: \left\{
\begin{array}{lllllll}
x'=3 (\frac{(x+y \theta \sqrt[3]{2})^{k+1}-(x-y \theta
\sqrt[3]{2})^{k+1}}{2 \sqrt[3]{2}(k+1)y \theta} - \frac{(x+y
\theta \sqrt[3]{2})^{k+2}+(x-y \theta
\sqrt[3]{2})^{k+2}}{\sqrt[3]{4}(k+1)(k+2)y^2 \theta^2} + \\
[3mm] + \frac{(x+y \theta \sqrt[3]{2})^{k+3}-(x-y \theta
\sqrt[3]{2})^{k+3}}{2(k+1)(k+2)(k+3)y^3 \theta^3} )$$
\\ [5 mm]$$y'=3(\frac{(x+y \theta \sqrt[3]{2})^{k+1}+(x-y \theta
\sqrt[3]{2})^{k+1}}{2 \sqrt[3]{4}(k+1)y \theta}- \frac{3((x+y
\theta \sqrt[3]{2})^{k+2}-(x-y \theta
\sqrt[3]{2})^{k+2})}{4(k+1)(k+2) y^2 \theta^2}+
\\ [3 mm]+\frac{3((x+y \theta \sqrt[3]{2})^{k+3}+(x-y \theta
\sqrt[3]{2})^{k+3})}{2 \sqrt[3]{2}(k+1)(k+2)(k+3)y^3 \theta^3} -
\frac{3((x+y \theta \sqrt[3]{2})^{k+4}-(x-y \theta
\sqrt[3]{2})^{k+4})}{2 \sqrt[3]{4}(k+1)(k+2)(k+3)(k+4)y^4
\theta^4})
\end{array}\right. \eqno(3.1)$$

\begin{pro} A function $\varphi \in C[0,1]$ is a solution of the Hammerstein's equation $$(H_kf)(t)=f(t)\eqno(3.2)$$
iff $\varphi(t)$ has the following form $$\varphi(t)=C_1 + C_2
\theta \sqrt[3]{4(t-\frac{1}{2})},$$ where $(C_1, C_2) \in R^2$ is
a fixed point of the operator $V_k$ (3.1).
\end{pro}
{\it Proof. Necessariness.} Let $\varphi \in C[0,1]$ be a solution
of the equation (3.2). Then we have $$\varphi(t)= C_1 + C_2 \theta
\sqrt[3]{4(t-\frac{1}{2})}, \eqno (3.3)$$ where
$$C_1=\int \limits_0^1 \varphi^k(u) \ du, \eqno(3.4)$$
$$C_2=\int \limits_0^1 \sqrt[3]{(u-\frac{1}{2})}\varphi^k (u)du \eqno(3.5) $$

Substituting function $\varphi(t)$ (3.3) into (3.4) we obtain

$$C_1=3 (\frac{(C_1+C_2 \theta \sqrt[3]{2})^{k+1}-(C_1-C_2
\theta \sqrt[3]{2})^{k+1}}{2 \sqrt[3]{2}(k+1)C_2 \theta} -
\frac{(C_1+C_2 \theta \sqrt[3]{2})^{k+2}+(C_1-C_2 \theta
\sqrt[3]{2})^{k+2}}{\sqrt[3]{4}(k+1)(k+2)C^2_2 \theta^2} + $$$$ +
\frac{(C_1+C_2 \theta \sqrt[3]{2})^{k+3}-(C_1-C_2 \theta
\sqrt[3]{2})^{k+3}}{2(k+1)(k+2)(k+3)C^3_2 \theta^3} ).$$

Substituting the function $\varphi(t)$ (3.3) into (3.5) we get

$$C_2=3(\frac{(C_1+C_2 \theta \sqrt[3]{2})^{k+1}+(C_1-C_2
\theta \sqrt[3]{2})^{k+1}}{2 \sqrt[3]{4}(k+1)C_2 \theta}-
\frac{3((C_1+C_2 \theta \sqrt[3]{2})^{k+2}-(C_1-C_2 \theta
\sqrt[3]{2})^{k+2})}{4(k+1)(k+2) C^2_2 \theta^2}+$$
$$+\frac{3((C_1+C_2 \theta \sqrt[3]{2})^{k+3}+(C_1-C_2 \theta
\sqrt[3]{2})^{k+3})}{2 \sqrt[3]{2}(k+1)(k+2)(k+3)C^3_2 \theta^3} -
\frac{3((C_1+C_2 \theta \sqrt[3]{2})^{k+4}-(C_1-C_2 \theta
\sqrt[3]{2})^{k+4})}{2 \sqrt[3]{4}(k+1)(k+2)(k+3)(k+4)C^4_2
\theta^4}).$$

Consequently the point $(C_1,C_2) \in R^2$ is a fixed point of the
operator $V_k$ (3.1).

{\it Sufficiency.} Suppose that a point $(C_1, C_2) \in R^2$ is a
fixed point of the operator $V_k$ define the function $\varphi(t)
\in C[0,1]$ by the equality
$$\varphi (t)=C_1+C_2 \theta \sqrt[3]{4(t-\frac{1}{2})}.$$

Then
$$(H_k \varphi) (t)= \int \limits_0^1 \left(1+\sqrt[3]{4} \theta \sqrt[3]{(t-\frac{1}{2})(u-\frac{1}{2})}\right) \varphi^k(u)du=
\int \limits_0^1 \varphi^k (u) du + $$ $$+\sqrt[3]{4} \theta
\sqrt[3]{t-\frac{1}{2}} \int \limits_0^1 \sqrt[3]{u-\frac{1}{2}}
\varphi^k(u) du=\int \limits_0^1 \left(C_1+C_2 \theta
\sqrt[3]{4(u-\frac{1}{2})} \right)^k du+$$$$+\sqrt[3]{4} \theta
\sqrt[3]{t-\frac{1}{2}} \int \limits_0^1
\sqrt[3]{u-\frac{1}{2}}\left(C_1+C_2 \theta
\sqrt[3]{4(u-\frac{1}{2})} \right)^k du =$$ $$3
\left(\frac{(\alpha+C_2 \theta \sqrt[3]{2})^{k+1}-(C_1-C_2 \theta
\sqrt[3]{2})^{k+1}}{2 \sqrt[3]{2}(k+1)C_2 \theta} - \frac{(C_1+C_2
\theta \sqrt[3]{2})^{k+2}+(C_1-C_2 \theta
\sqrt[3]{2})^{k+2}}{\sqrt[3]{4}(k+1)(k+2)C^2_2 \theta^2}\right.+$$
$$\left.+\frac{(C_1+C_2 \theta \sqrt[3]{2})^{k+3}-(C_1-C_2 \theta
\sqrt[3]{2})^{k+3}}{2(k+1)(k+2)(k+3)C^3_2
\theta^3}\right)+3\sqrt[3]{4} \theta \sqrt[3]{t-\frac{1}{2}}
\times$$
$$ \times \left(\frac{(C_1+C_2 \theta \sqrt[3]{2})^{k+1}+(C_1-C_2 \theta
\sqrt[3]{2})^{k+1}}{2 \sqrt[3]{4}(k+1)C_2 \theta}-
\frac{3((C_1+C_2 \theta \sqrt[3]{2})^{k+2}-(C_1-C_2 \theta
\sqrt[3]{2})^{k+2})}{4(k+1)(k+2) C^2_2 \theta^2}+\right.$$
$$\left.\frac{3((C_1+C_2 \theta \sqrt[3]{2})^{k+3}+(C_1-C_2 \theta
\sqrt[3]{2})^{k+3})}{2 \sqrt[3]{2}(k+1)(k+2)(k+3)C^3_2 \theta^3} -
\frac{3((C_1+C_2 \theta \sqrt[3]{2})^{k+4}-(C_1-C_2 \theta
\sqrt[3]{2})^{k+4})}{2 \sqrt[3]{4}(k+1)(k+2)(k+3)(k+4)C^4_2
\theta^4}\right)=$$ $=C_1+C_2 \theta
\sqrt[3]{4(t-\frac{1}{2})}=\varphi(t),$

i.e. the function $\varphi(t)$ is a solution of the equation
(3.2).

\section{A phase transition for the model (2.1) at $k=2$}

For $k=2$ the operator
$V_{2}:(x,y)\in\mathbb{R}^{2}\rightarrow(x',y')\in\mathbb{R}^{2}$
(see (3.1)) has the form

$$\left\{
\begin{array}{cc}
  x'=x^{2}+\frac{3\sqrt[3]{4}}{5}\theta^{2}y^{2}, \\ [2 mm]
  y'=\frac{6}{5}\theta x y. \\
\end{array}\right. \eqno(4.1)
$$

\begin{pro}\label{p1} a) If $0\leq\theta\leq\frac{5}{6},$ then the Hammerstein's operator $H_{2}$ has unique
(nontrivial) positive fixed point in the $C[0,1]$. \\
b) If $\frac{5}{6}<\theta<1,$ then there are exactly three
positive fixed points in $C[0,1]$ of the Hammerstein's operator.
\end{pro}
\proof Clearly, that in the case $\theta=0$ the Hammerstein's
operator $H_{2}$ has unique nontrivial positive fixed point
$\varphi(t)\equiv1$. Let $\theta\neq0$. We consider the system of
equations for a fixed point of the operator $V_{2}:$

$$ \left\{
\begin{array}{ccc}
  x^{2}+\frac{3\sqrt[3]{4}}{5}\theta^{2}y^{2}=x, \\ [2 mm]
  \frac{6}{5} \theta xy=y.
\end{array} \right. \eqno (4.2)
$$

In the case $y=0$ from (4.2) we have two solutions (0;0) and
(1;0). By proposition 3.1 functions
$$\varphi(t)=\varphi_{0}(t)\equiv0, \,\
\varphi(t)=\varphi_{1}(t)\equiv 1 $$ are solutions of the equation
(4.2).

Suppose $y\neq0$ in the (4.2). Then from (4.2) we get
$x=\frac{5}{6\theta}.$ Consequently, from the first equation of
(4.6) we get
$$y^{2}=\frac{25}{3\sqrt[3]{4}\theta^{2}}\cdot\frac{6\theta-5}{36\theta^{2}}.$$
Hence it follows, that $\theta>\frac{5}{6}$ and
$$y=y^{\pm}_{1}=\pm\frac{5}{6\theta^{2}}\cdot\frac{1}{\sqrt[3]{2}}\cdot\sqrt{\frac{6\theta-5}{3}}.$$
Thus, in the case $0\leq\theta\leq\frac{5}{6}$ operator $V_{2}$
has two fixed points: (0;0), (1;0) and in the case
$\frac{5}{6}<\theta<1$ the operator $V_{2}$ has four fixed points:
(0;0), (1;0), $(x_{1},y^{+}_{1})$ and $(x_{1},y^{-}_{1}),$ with
$x_{1}=\frac{5}{6\theta}.$

Note that, there is no any other fixed point for $V_{2}$.

Consequently,  $$\varphi_{1}(t)\equiv1,$$
$$\varphi_{2}(t)=\frac{5}{6\theta}\left(1+\sqrt{\frac{6\theta-5}{3}}\cdot\sqrt[3]{2\left(t-\frac{1}{2}\right)}\right),$$
$$\varphi_{3}(t)=\frac{5}{6\theta}\left(1-\sqrt{\frac{6\theta-5}{3}}\cdot\sqrt[3]{2\left(t-\frac{1}{2}\right)}\right)$$
are non trivial fixed points of the Hammerstein's operator
$H_{2}.$ Thus we have proved the following

\begin{thm}\label{t4.2.} a) If $ 0\leq\theta\leq\frac{5}{6},$ then  for the model (2.1) on the Cayley
tree $\Gamma^{2}$ there exists a unique translational -- invariant Gibbs measure; \\
b) If $\frac{5}{6}<\theta<1,$ then for the model (2.1) on the
Cayley tree $\Gamma^{2}$ there are three translational --
invariant Gibbs measures.
\end{thm}

\section{A phase transition for the model (2.1) at $k=3$}

For $k=3$ the operator
$V_{3}:(x,y)\in\mathbb{R}^{2}\rightarrow(x',y')\in\mathbb{R}^{2}$
(see (3.1)) has the form

$$ \left\{
\begin{array}{cc}
  x'=x^{3}+\frac{18}{5}\cdot\frac{\theta^{2}}{\sqrt[3]{2}}x y^{2},
  \\ [2 mm]
  y'=\frac{9}{5}\theta x^{2} y + \frac{6}{7}\cdot\frac{\theta^{3}}{\sqrt[3]{2}}y^{3}. \\
\end{array} \right.\eqno(5.1)
$$

\begin{pro}\label{p1} a) If $0\leq\theta\leq\frac{5}{9},$ then (in
the $C[0,1]$) Hammerstein's operator $H_{3}$ has a unique
nontrivial
positive fixed point; \\
b) If $\frac{5}{9}<\theta<1,$ then there are exactly three
positive fixed points of the Hammerstein's operator $H_{3}$
$C[0,1]$.
\end{pro}

\proof Clearly, that in the case $\theta=0$ operator $H_{3}$ has
unique positive fixed point $\varphi(t)\equiv1$. Let
$\theta\neq0$. We consider

$$ \left\{
\begin{array}{cc}
  x^{3}+\frac{18}{5}\cdot\frac{\theta^{2}}{\sqrt[3]{2}}x y^{2}=x,
  \\ [2 mm]
  \frac{9}{5}\theta x^{2} y + \frac{6}{7}\cdot\frac{\theta^{3}}{\sqrt[3]{2}}y^{3}=y. \\
\end{array} \right. \eqno(5.2)
$$

For $y=0$ from (5.2) we have three solutions of the system
equations (5.2): (0;0), (-1;0) and (1;0). By proposition 3.1
functions
$$\varphi(t)=\varphi_{0}(t)\equiv0, \,\
\varphi(t)=\varphi_{1}^{\pm}(t)\equiv\pm1 $$ are solutions of the
equation $H_3f=f$. For $x=0$ from (5.2) we get three solutions
$(0;0), (0;y^{+}_{1}),$ $ (0;y^{-}_{1}),$ where
$$y^{\pm}_{1}=\pm\frac{\sqrt[6]{2}}{\theta}\cdot\sqrt{\frac{7}{6\theta}}.$$
Hance the functions
$$\varphi_{2}^{\pm}(t)=\pm\sqrt[6]{2}\cdot\sqrt{\frac{7}{6\theta}}\cdot\sqrt[3]{4\left(t-\frac{1}{2}\right)}$$
are solutions of the equation $H_3f=f$.

Suppose that $x\neq0$ and $y\neq0$ in the (5.2). Then the system
of equations (5.2) can be rewritten as

$$ \left\{
\begin{array}{cc}
  x^{2}+\frac{18}{5}\cdot\frac{\theta^{2}}{\sqrt[3]{2}} y^{2}=1,
  \\ [2 mm]
  \frac{9}{5}\theta x^{2} + \frac{6}{7}\cdot\frac{\theta^{3}}{\sqrt[3]{2}}y^{2}=1. \\
\end{array} \right.  \eqno(5.7)
$$

Hence it follows
$$x^{2}=1-\frac{18}{5}\cdot\frac{\theta^{2}}{\sqrt[3]{2}}y^{2}$$
and
$$\frac{9}{5}\theta\left(1-\frac{18}{5}\cdot\frac{\theta^{2}}{\sqrt[3]{2}}y^{2}\right)+\frac{6}{7}\cdot\frac{\theta^{3}}{\sqrt[3]{2}}y^{2}=1.$$
Then
$$y^{2}=\frac{105}{164}\cdot\frac{\sqrt[3]{2}}{2\theta^{2}}\cdot\frac{9\theta-5}{9\theta}.$$
Therefore $\theta>\frac{5}{9}$ and
$$y=y^{\pm}_{2}=\pm\sqrt{\frac{105}{164}}\cdot\frac{1}{\theta\sqrt[3]{2}}\cdot\sqrt{\frac{9\theta-5}{9\theta}}.$$
Consequently
$$x^{2}=1-\frac{21}{164}\cdot\frac{9\theta-5}{\theta}.$$

Hence we conclude that

$$x=x^{\pm}_{1}=\pm\sqrt{1-\frac{21}{164}\cdot\frac{9\theta-5}{\theta}}.$$

Thus, the operator $V_{3}$ (5.1) has five fixed points: $(0;0),
\,\ (-1;0), \,\ (1;0), \,\ (0;y^{+}_{1})$ and $(0;y^{-}_{1})$, if
$0\leq\theta\leq\frac{5}{9}$ and $V_{3}$ has nine fixed points:
$(0;0), \,\ (-1;0), \,\ (1;0),$ $(0,y^{+}_{1}),$ $(0;y^{-}_{1})$
$(x^{+}_{1};y^{+}_{1}),$ $(x^{+}_{1};y^{-}_{1}),$
$(x^{-}_{1};y^{+}_{1})$ and $(x^{-}_{1};y^{-}_{1})$, if
$\frac{5}{9}<\theta<1.$

Note that the above mentioned solutions are all possible solutions
of $V_{3}$.

Consequently by proposition 3.1 the operator $H_{3}$ has unique
positive fixed point $\varphi(t)=\varphi_{1}(t)\equiv1$ if
$0\leq\theta\leq\frac{5}{9}.$ In the case $\frac{5}{9}<\theta<1$
the functions
$$\varphi_{1}(t)\equiv1, \,\ \varphi_{2}(t)=x^{+}_{1}+y^{+}_{1}\theta\sqrt[3]{4\left(t-\frac{1}{2}\right)},
\,\
\varphi_{3}(t)=x^{+}_{1}+y^{-}_{1}\theta\sqrt[3]{4\left(t-\frac{1}{2}\right)}$$
are positive fixed points of the Hammerstein's operator
$H_{3}.$\endproof

From Proposition 3.1 and Proposition 5.1 it follows that

\begin{thm}\label{t4.2.} a) If $\ 0\leq\theta\leq\frac{5}{9}$ for the model (2.1) on the Cayley tree
$\Gamma^{3}, $ then there exists a
uniqie translational -- invariant Gibbs measure;\\
b) If $\frac{5}{9}<\theta<1,$ then there exist there translational
-- invariant Gibbs measures.
\end{thm}

\end{document}